\documentclass[twoside]{article}

\usepackage{latexsym}      
\usepackage{graphicx}      
\usepackage{amsmath}
\usepackage{braket}

\usepackage[margin=1.0in,papersize={21.0cm,29.7cm}]{geometry}

\newcommand{\changefont}{%
    \fontsize{9}{10}\selectfont
}

\usepackage{fancyhdr}
\tiny
\makeatletter
\renewcommand\section{\@startsection
   {section}{1}{0pt}%
   {-\baselineskip}%
   {0.1\baselineskip}%
   {\normalfont\large\bfseries}}%
\makeatother

\usepackage{caption}
\usepackage[compress,nospace]{cite}

\begin{document}

\begin{center}
{\Large\textbf{Subluminal velocity of OAM-carrying beam}}\\
\vspace{0.05in}
\textbf{Nestor D. Bareza Jr.$^{\ast}$ and Nathaniel Hermosa}\\
\textit{National Institute of Physics, University of the Philippines Diliman, Quezon City, Diliman 1101}\\
\textrm{$^{\ast}$Corresponding author: nbareza@nip.upd.edu.ph}\\

\vspace{0.15in}

\parbox{4.5in}{{\large \textbf{Abstract}}\\
\noindent{We report a consequence of the orbital angular momentum (OAM) of a beam to its group velocity. We calculate the group velocity $v_g$ of Laguerre-Gauss beam (\emph{LG}) with $\ell$ and at $p=0$. The $v_g$ reduction of \emph{LG} beam even in free space is observed to have dependence on both orbital or winding number $\ell$ and the beam's divergence $\theta_0$. We found that light possessing higher $\ell$ travels relatively slower than that with lower $\ell$ values. This suggests that light of different OAM separate in the temporal domain along propagation and it is an added effect to the dispersion due to field confinement. Our results are useful for treating information embedded in light with OAM from astronomical  sources and/or data transmission in free space.}\\

}\\
\end{center}

The speed of light \emph{c}, is constant. This fact has been a pillar of modern physics since the derivation of Maxwell and in Einstein's postulate in Special Relativity, and it has been a basic assumption in light's various applications. A physical beam of light however, has a finite extent such that even in free space it is by nature dispersive. The confinement of the field changes its transverse wave vector, hence, altering the light's group velocity $v_g$. Confined light therefore, would have its $v_g$ that is not equal to \emph{c}.

Recently, Giovannini et al showed thru experiments, backed by calculations, that spatially structured light indeed travels slower than \emph{c} \cite{giovanninispatially}. That is, there is a decrease of  group velocity for structured light. Although the phenomenon can be explained classically, they used a Hong-Ou-Mandel interferometer to measure the lag of a structured photon compared to a photon which was not structured. In their experiment, the slowing of light is due to dispersion in free space. They performed their experiment with a Bessel beam and a Gaussian beam. Alfano and Nolan remarked that by considering dispersion relation, Bessel beam can be very slow near a critical frequency which can be used as optical buffer in free space \cite{alfanoslowing}. Slowing light due to its structure is different from slowing light with materials.

\begin{figure}[h!]
\centering
\includegraphics[width=4in]{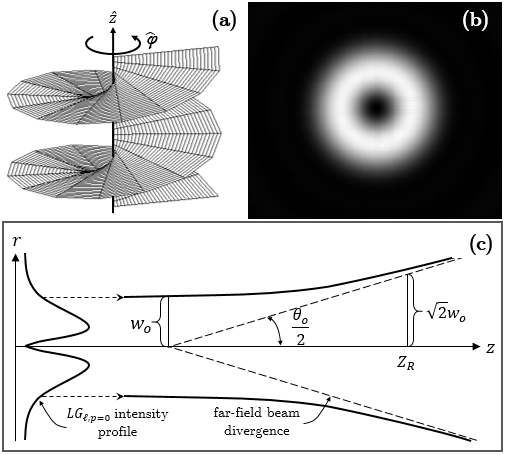}
\caption{Illustrations of (a) \emph{LG} wavefront for $\ell$=1 with its corresponding (b) intensity profile and (c) \emph{LG} beam spreading through propagation.}
\label{fig:LG}
\end{figure}

A structured light that has recently gained much attention are beams endowed with orbital angular momentum (OAM). It has a phase factor of the form $\ell\varphi$, where $\varphi$ is the azimuthal angle and $\ell$ is the number of $2\pi$ windings around $\varphi$. First investigated by Allen et al, these beams have Poynting vectors that spiral along the direction of propagation \cite{allenorbital}.  The skewness of the Poynting vector with respect to optical axis $z$ is inversely related to radial distance $r$ \cite{leachdirect}. The wavefront for this kind of beam which is helical is illustrated in Figure \ref{fig:LG}$a$. The negative $\ell$s will yield the same wavefronts but of opposite helicities. The intensity profile of this kind of beam is a donut shape with smooth radial gradients as shown in Figure \ref{fig:LG}$b$. Each photon in such beams carries an OAM value of $\ell\hbar$.  This beam's property has led to a myriad of applications from optical tweezing and micromanipulation \cite{curtisdynamic, galajarotational}, to free space information \cite{gibsonfreespace}, to tranverse Doppler effect \cite{rosalesexperimental}, and in astrophysics \cite{berkhoutmethod}. In the paraxial regime, these beams form a complete basis set such that it can be used as a tool in quantum information processes \cite{mairentanglement, dambrosioorbital, molinatwisted}.

It is therefore important to ask: \emph{What is the effect of the OAM of a beam on its group velocity?} The consequences are extensive. The most important of which is the different time of arrival of information even in free space propagation. This is similar to modal dispersion in fiber, a serious limitation in optical fiber communication. The promised massive information when using beams with OAM will have an issue. Information embedded in these beams will not arrive at the same time and some corrections are then necessary. 

In this letter, we report our calculation on the reduction of $v_g$'s due to the beam's OAM. The analytical expression is exact with just a factor $\ell$ and our expression reduces to the results of Giovannini when $\ell=0$. 

We follow the derivation in \cite{giovanninispatially}, with a Laguerre-Gauss (\emph{LG}) beam in the lowest radial order $p=0$ which is given by
\begin{equation}
\label{eqn:LG}
\psi_{l,p=0}(r, \varphi) \sim r^{|\ell|}exp\left(-\frac{r}{w(z)}\right)^2exp\left(-i\ell\varphi\right)exp\left(\frac{-ikr^2}{R(z)}\right),
\end{equation}

\noindent where $w(z)=w_o\sqrt{1+\frac{z^2}{z_R^2}}$, $R(z)=\frac{z^2+Z_R^2}{z}$ and $Z_R=\frac{1}{2}kw_o^2$ is the Rayleigh length. This beam spreads along propagation as illustrated in Figure \ref{fig:LG}$c$. In the far-field, the divergence due to Gaussian term which is represented by opening angle $\theta_o$ can be simply expressed in terms of minimum beam waist $w_o$ and wavevector $k_o$:
\begin{equation}
\label{eqn:diverge}
\frac{\theta_o}{2}=\frac{2}{k_ow_o}.
\end{equation}

\captionsetup[figure]{width=6in}

The effective group velocity of a beam propagating along $z$ and with central frequency $\omega_0$ is 

\begin{equation}
\label{eqn:v_g}
v_g^{(z)}=\left(\frac{\partial^2\Phi}{\partial z \partial \omega}\right)^{-1}_{\omega_0},
\end{equation}

\noindent where $\Phi$ is the phase profile. $\Phi$ can be represented as the complex argument of functional value of $\psi$, which is the transverse profile in a scalar approximation as in \cite{giovanninispatially}:

\begin{equation}
\label{eqn:psi}
\Phi=arg\left<\psi(z,\omega_o)|\psi(z,\omega)\right>
\end{equation}

Fields confined within a volume would have an effective $v_g$ as the harmonic mean of the velocity and weighted by the local intensity. The average $v_g$ for path interval $\Delta z=z_2-z_1$ is then given by, 

\begin{equation}
\label{eqn:ave_v}
v_g=\left(\frac{1}{z_2-z_1}\int^{z_2}_{z_1}\left(\frac{\partial^2\Phi}{\partial z\partial\omega}\right)_{\omega_0}dz\right)^{-1}=\Delta z/\left[\frac{\partial}{\partial\omega}\left(arg\left<\psi\left(z,\omega_0\right)|\psi\left(z,\omega\right)\right>\right)_{\omega_0}\right]^{z_2}_{z_1}.
\end{equation}

\noindent The denominator in the rightmost side denotes the time $\Delta t$ it takes for the structured light to travel $\Delta z$ with an added path length  $\delta z$ due to its altered wavevector that also contains some transverse components. They are related by $\Delta t= \left(\Delta z+\delta z\right)/\emph{c}$. The added path length can then be expressed in terms of complex argument of the functional value of $\psi$ in wavevector domain with $k_o$ as the central wavevector:

\begin{equation}
\label{eqn:delz}
\delta z=\left[\frac{\partial}{\partial k}\left(arg\left<\psi\left(z,k_0\right)|\psi\left(z,k\right)\right>\right)_{\omega_0}\right]^{z_2}_{z_1}-\Delta z.
\end{equation}

Now, $v_g$ can be reformulated in terms of $\delta z$ given in (\ref{eqn:delz}) by the following,

\begin{equation}
\label{eqn:vgdelz}
v_g=c/\left(1+\frac{\delta z}{\Delta z}\right).
\end{equation}

\noindent Translating the spatial dependence of this $v_g$ into wavevector, we solve (\ref{eqn:delz}) in the paraxial regime whose wave equation is 

\begin{equation}
\label{eqn:parax}
\frac{\partial\psi}{\partial z}=\frac{i}{2k}\nabla_{\perp}^2\psi+ik\psi
\end{equation}

\noindent where $\nabla_{\perp}^2$ operator is the transverse laplacian, which can also be written in terms of quantum mechanical operator to transverse wavevector $\hat{k_{\perp}}=-i\nabla_{\perp}$. In this regime, the wavefunction evolves from $z_1$ to $z_2$ by $\ket{\psi(z_2,k)}=\text{exp}\left[\Delta z\left(\frac{i}{2k}\nabla_{\perp}^2+ik\right)\right]\ket{z_1,k}$. By taking an inner product of $\psi(z_2,k)$ and subsituting result to (\ref{eqn:delz}), we obtain the relations:

\begin{equation}
\label{eqn:zk}
\frac{\delta z}{\Delta z}=\frac{\braket{\hat{k}_{\perp}^2}_{\ket{\psi(z_1, k_o)}}}{2k_o^2}
\end{equation}

\noindent so that (\ref{eqn:ave_v}) becomes,
\begin{equation}
\label{eqn:vg}
v_g=\frac{c}{1+\frac{\left<k_{\perp}^2\right>}{2k_o^2}}
\end{equation}

\noindent where,
\begin{equation}
\label{eqn:k2}
\left<k_{\perp}^2\right> = \frac{\left<\psi|k_{\perp}^2|\psi\right>}{\left<\psi|\psi\right>}.
\end{equation}

\noindent Using \emph{LG} beam with $\ell$ and at $p=0$ as the wavefunction, we substitute ($\ref{eqn:LG}$) to ($\ref{eqn:k2}$) in order to have 

\begin{equation}
\label{eqn:k2LG}
\left<k_{\perp}^2\right>= \frac{2}{w_0^2}(|\ell|+1).
\end{equation}

\noindent And the group velocity for such beam is given by,

\begin{equation}
\label{eqn:v_g_l}
v_g=\frac{c}{1+\left(\frac{1}{k_0w_0}\right)^2\left(|\ell|+1\right)}=\frac{c}{1+\left(\frac{\theta_0}{4}\right)^2\left(|\ell|+1\right)}
\end{equation}

\noindent where $1/k_0w_0$ is replaced by the opening angle of the beam $\theta_0/4$ for a more intuitive picture. The absolute value appearing in $\ell$ indicates that same $v_g$ reduction is obtained regardless of $\ell$ polarity or wavefront helicity. When $\ell=0$, $v_g = c/\left(1+\frac{1}{k_0^2w_0^2}\right)$ which is consistent with the calculation for Gaussian beam in \cite{giovanninispatially}. 

Eqn. (\ref{eqn:v_g_l}) is the most important result in this letter. This consequently shows that an \emph{LG} beam is subluminal and dispersive even in free-space. The $v_g$ reduction depends on both orbital order $\ell$ and far field beam divergence $\theta_o$. For light with OAM, we can think that the added path length due to beam divergence increases by a factor of $|\ell|+1$. This factor is consistent to the conservation of total linear momentum in the system. Moreover, this factor amplifies the time delay for light to travel the path difference $\Delta z$ hence, a lower $v_g$ occurs. In the work of Giovannini et al \cite{giovanninispatially}, the added path length comes from the radial component of Poynting vector with respect to the optical axis. In equation (\ref{eqn:k2LG}), we show that even a Poynting vector with angular component with respect to the optical axis can also contribute to the path.

\begin{figure}[h!]
\centering
\includegraphics[width=6in]{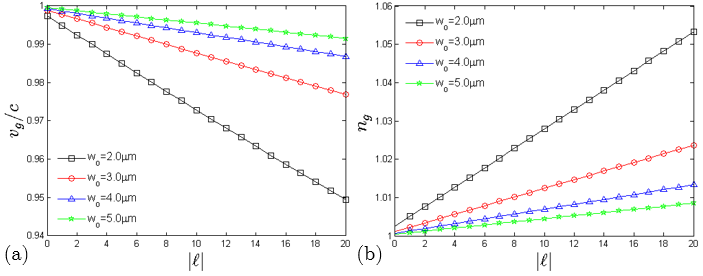}
\caption{Plots of (a) $v_g$/\emph{c} and (b) $n_g$ versus $|\ell|$ for different minimum beam waist $w_o$ values.}
\label{fig:vgng}
\end{figure}

The $v_g$ reduction is depicted by displaying plots between the ratio $v_g/\emph{c}$ versus $\ell$ for different $w_o$ values, with the beam's central wavelength set to $\lambda_0=632.8nm$, as shown in Figure \ref{fig:vgng}(a). All $v_g/c$ values obtained are lower than unity implying the subluminal behavior of \emph{LG} beams. Notice first that even for $\ell=0$, the subluminality of $v_g$ is apparent for different $w_o$ values, and that $v_g$ is even more reduced for relatively smaller $w_o$. This holds true since, for a certain $\lambda_o$, relatively lower $w_o$ yields larger far-field beam divergence. As the beam propagates for such case, the field confinement in the transverse structure is amplified. Now as $\ell$ increases, $v_g$ becomes even slower, which can be observed for different $w_o$ values. This indicates that a beam with higher orbital order will have greater path length $\delta z$, which is evident when relating (\ref{eqn:k2LG}) to (\ref{eqn:zk}). A simple calculation based on (\ref{eqn:v_g_l}) gives an almost 5\% speed reduction for a beam with $w_0= 2\mu m$ and $\ell=20$.
 
Furthermore, eqn. (\ref{eqn:v_g_l}) shows that light-carrying OAM is experiencing free-space dispersion. The effective group index of refraction $n_g$ is given by,
\begin{equation}
n_g=1+\left(\frac{\theta_0}{4}\right)^2\left(|\ell|+1\right)
\end{equation}

\noindent It can be observed from plots displayed in Figure \ref{fig:vgng}(b) that for any $w_o$ values, $n_g$ is linearly related to $|\ell|$. Thus, beams of different $\ell$ values that are initially propagated simultaneously will have different time delays after travelling the same $\Delta z$. This makes \emph{LG} beams separate in the temporal domain which consequently demands corrections in its applications such as in data transmission/communication, in multiplexing, and in OAM spectrum detection \cite{wangterabit, karimitime, tamburiniencoding, laveryefficient}.

\captionsetup[figure]{width=6in}

Padgett et al. demonstrate that for a given beam size, the far-field opening angle increases with increasing OAM \cite{padgettdivergence}. This consequently suggests that a larger aperture is required when receiving beams with relatively higher OAM. The $\ell$-dependence of $v_g$ for \emph{LG} beams that we reported may be incorporated to such receiving optical system. A time-controllable receiving aperture size can be programmed according to computed delays prior to the arrival of beams. As opposed to the beam divergence relation presented in (\ref{eqn:diverge}) which is due to skewness of Poynting vector with respect to optical axis, they also considered the contribution of normal diffractive spreading by the standard deviation of the spatial distribution. They derived the far-field beam divergence $\alpha_{\ell}$ to be dependent on $|\ell|$ in the following manner,
\begin{equation}
\label{eqn:alpha}
\alpha_{\ell}=\sqrt{\frac{|\ell|+1}{2}}\frac{2}{k_ow_o}
\end{equation}
\noindent Reformulating the calculated $v_g$ for LG beams according to this beam divergence definition, we get a more compact form given by
\begin{equation}
\label{eqn:vgalpha}
v_g=\frac{c}{1+\frac{\alpha_{\ell}^2}{2}}
\end{equation}

In an actual experiment where a Gaussian laser beam traversed a fork hologram, the intensity profile deviates from the \emph{LG} expression given by eqn. (\ref{eqn:LG}) and is expressed as Kummer function. Kummer beams yielded by confluent hypergeometric function, are expressed as the difference between two Bessel functions \cite{ricciinstability}. The wavefunction for such case is extensively derived in \cite{janicijevicfresnel} in which the model is more apparent in the far-field region. This is however beyond the scope of this letter.

In conclusion, we have derived the group velocity $v_g$ of \emph{LG} beam that depends on both orbital order $|\ell|$ and far-field beam divergence $\theta_o$. This result shows that light is both subluminal and dispersive even in free space. This would have far-reaching consequences on the OAM beam's applications.

\end{document}